\documentclass[aps,prd,twocolumn,showpacs,superscriptaddress,10pt]{revtex4-1}

\usepackage{acronym}
\usepackage{amsfonts,amsmath,amssymb}
\usepackage{bm}
\usepackage{cases}
\usepackage{comment}
\usepackage[T1]{fontenc} 
\usepackage{graphicx}
\usepackage{indentfirst}
\usepackage{mathrsfs}
\usepackage{multirow}
\usepackage{scrextend}
\usepackage{hyperref}
\usepackage{xcolor}
\usepackage{soul}
\usepackage[most]{tcolorbox}

\newcommand{\jeven}{\mbox{\rm\j}}

\begin{document}

\title{Master functions of Reissner-Nordstr\"om black hole perturbations and their Darboux transformation}

\author{Hui-Fa Liu}
\email{hfaliu@163.com}
\affiliation{Beijing University of Technology, Beijing 100124, China}

\author{Ding-fang Zeng}
\email{dfzeng@bjut.edu.cn}
\affiliation{Beijing University of Technology, Beijing 100124, China}

\date{\today}

\begin{abstract}

Lenzi and Sopuerta developed a new method to construct master functions for the perturbation of vacuum black holes. We extend this method to black holes coupled with electromagnetic field and cosmological constant by allowing the master functions to be linear combinations of the metric and electromagnetic-potential perturbations, as well as their first-order derivatives. Requiring these master functions satisfy wave equations with yet-to-be-determined effective potentials, we reduce the linearized Einstein--Maxwell system to a set of algebraic-differential constraints. Solving these constraints reveals four master function branches in each parity sector: two standard branches, which coincide with the Zerilli-Moncrief formalism, and two Darboux branches, characterized by their effective potentials. Within each parity sector, a Darboux transformation exists which connects the standard and Darboux branches, preserving the quasinormal mode spectrum and confirming their physical equivalence.

\end{abstract}

\maketitle

\section{Introduction}

The gravitational waves predicted by general relativity were first directly detected in 2015 by the LIGO collaboration in a binary black hole merger \cite{LIGOScientific:2016aoc}. This landmark discovery opened a new observational window for exploring strong-field spacetime geometry. With the enhanced sensitivity of detectors in the LIGO-Virgo-KAGRA collaboration \cite{LIGOScientific:2021tfm,LIGOScientific:2021qlt} and the upcoming launch of the space-based interferometer LISA, which is designed to capture millihertz gravitational waves \cite{LISA:2017pwj,LISA:2022yao}, developing accurate waveform models has become increasingly critical. These models are essential for extracting the mass, spin, and other dynamical parameters of the merging bodies, from the gravitational-wave signals. Such advances are central not only to tests of general relativity in the strong-field regime but also to the exploration of the formation and evolution of black holes.

Accurate extraction of strong-field geometric information from the gravitational-wave signals relies on detailed modeling of black hole dynamics. Analytical solutions to the Einstein equations, such as the Schwarzschild, Kerr, Reissner-Nordstr\"om (RN), and Kerr-Newman metrics, provide a baseline description for static or stationary black holes but do not reflect dynamics of the extreme-mass-ratio inspirals \cite{Pound:2021qin} or the ringdown phase of binary-merger systems. In contrast, black hole perturbation theory offers a valuable description for these systems through their quasinormal modes (QNMs), which are damped oscillatory modes whose eigenfrequencies are determined by the master equation \cite{Chandrasekhar:1983} (for reviews, see \cite{Nollert:1999ji,Kokkotas:1999bd,Berti:2009kk,Konoplya:2011qq,Bolokhov:2025uxz}). These modes carry information of the black hole parameters and can be measured through gravitational wave spectroscopy techniques \cite{Cotesta:2022pci,Farr:2013yna,Ghosh:2024het}.

Master equations are key ingredients of black hole perturbation theory. For nonrotating black holes, Regge and Wheeler established the framework of spherical harmonic decomposition of the linearized Einstein equations, separating them into odd-parity and even-parity modes. Their seminal work yielded the master equation for odd-parity perturbations of Schwarzschild geometry \cite{Regge:1957td}. Several years later, Zerilli introduced a decoupling procedure and obtained the master equation for even-parity perturbations \cite{Zerilli:1970se,Zerilli:1970wzz}, see also \cite{Moncrief:1974am}. These works were extended to the charged black holes \cite{Zerilli:1974ai,Moncrief:1974gw,Moncrief:1974ng}, modified-gravity scenarios \cite{Liu:2023uft} and higher-dimensional spacetimes in Refs.~\cite{Kodama:2000fa,Kodama:2003jz,Ishibashi:2003ap,Kodama:2003kk}. For rotating black holes, by the Newman-Penrose formalism \cite{Newman:1961qr}, Teukolsky obtained separable equations \cite{Teukolsky:1972my,Teukolsky:1973ha} for the Weyl scalar, a curvature representation of the metric perturbation. This method has become the standard for rotating black hole perturbations \cite{Loutrel:2020wbw,Ripley:2020xby,Spiers:2023cip}. 

These traditional approaches for obtaining the master equations rely on \emph{ad hoc} manipulations of the coupled field equations. Such manipulations usually yield a unique master function and associated equation for each parity sector. In Refs.~\cite{Lenzi:2021wpc,Lenzi:2024tgk}, Lenzi and Sopuerta proposed a new method which has the potential to obtain all possible master functions and equations for given parity sectors. For black holes in vacuum, their method yields master functions with two-branch features. In the odd-parity sector, besides the branch that contains the Regge-Wheeler and Cunningham-Price-Moncrief \cite{Cunningham:1978zfa,Cunningham:1979px,Cunningham:1980cp} master functions, there exists an additional branch which is defined by a master equation characterized by different effective potential. The even-parity sector exhibits similar two-branch structure. Since practical black holes always live in complex astronomical environment, their coupling with matter field is very natural. This raises questions: how will the coupling between the black hole and the environmental matter change the landscape of master functions? Can the two-branch structure persist when the matter field perturbation is included?

In this paper, we address these questions by extending Lenzi and Sopuerta's method to the perturbation of black holes coupled with electromagnetic field and cosmological constant. We focus on the construction of the master functions and their associated equations, in a general perturbative gauge. Our procedure is as follows. (i) We reduce the four-dimensional linearized Einstein--Maxwell equations --- using spherical harmonic decomposition --- to a system of two-dimensional partial differential equations. (ii) Within each parity sector, we express the master function as a linear combination of the metric and electromagnetic-potential perturbations and their first-order derivatives. We then require that the master functions satisfy wave equations with yet-to-be-determined effective potentials. (iii) We derive the resultant algebraic-differential constraints on both the coefficients of the master function ansatz and the effective potential from the reduced field equations. (iv) By solving these constraints, we obtain expressions for the master functions and their associated potentials. We find a four-branch structure in each parity sector, with two branches matching the Zerilli-Moncrief formalism \cite{Zerilli:1974ai,Moncrief:1974gw,Moncrief:1974ng}, whereas the other two have not been explored previously. Finally, we employ the Darboux transformation to reveal intrinsic relationships among the resultant master equations.

The remainder of this paper is organized as follows. In Sec. \ref{decomposition-of-black-hole-perturbations}, we derive the linearized Einstein--Maxwell equations and present the perturbation decomposition framework. Section \ref{construction-master-functions-and-equations} details the construction of the master functions and equations in a general perturbative gauge. In Sec. \ref{Darboux-transformation}, we explore the Darboux transformation connecting different master equations. Section \ref{conclusions} summarizes our calculation and key results. Throughout this paper, we adopt the natural units $c = 8\pi G = 1$ and the metric signature $(-,+,+,+)$.


\section{Decomposition of Black-Hole Perturbations}
\label{decomposition-of-black-hole-perturbations} 

Our starting point is a four-dimensional spherically symmetric black hole coupled with electromagnetic field and cosmological constant $\Lambda$. The Einstein--Maxwell equations have the form
\begin{gather}
R_{\mu\nu}-\frac{1}{2}(R-2\Lambda)g_{\mu\nu}=2\bigl(F_{\mu\rho}F_{\nu}^{~\rho}-\frac{1}{4}g_{\mu\nu}F_{\rho\sigma}F^{\rho\sigma}\bigr)\,,
\\
{F^{\mu\nu}}_{;\nu}=0 \,,\quad F_{\mu\nu}\equiv{}A_{\nu,\mu}-A_{\mu,\nu}\,.
\label{eq-Einstein–Maxwell}
\end{gather}
For the unperturbed (static) solution, the metric $g_{\mu\nu}$ and electromagnetic-potential $A_{\mu}$ are
\begin{gather}
g_{\mu\nu}dx^{\mu}dx^{\nu}=-f(r)dt^2+\frac{1}{f(r)}dr^2+r^2d\Omega^2 \,,
\label{ds2-g0}
\\
f(r)=1-\frac{2M}{r}-\frac{\Lambda}{3}r^2+\frac{Q^2}{r^2}\,,\quad A_{\mu}dx^{\mu}=-\frac{Q}{r}\,dt \,,
\label{unperturbed-A}
\end{gather}
where $d\Omega^2=d\theta^{2}+\sin^{2}\theta d\phi^{2}$ is the line element on the unit 2-sphere, while $M$ and $Q$ are the mass and charge of the black hole, respectively. For $\Lambda=0$, the solution reduces to the standard RN black hole, while for $\Lambda>0$ and $\Lambda<0$, it corresponds to the RN-de Sitter and RN-anti--de Sitter black holes, respectively.

We now introduce first-order perturbations \cite{Brizuela:2006ne}. We will write the perturbed metric and electromagnetic field as follows
\begin{eqnarray}
\tilde{g}_{\mu\nu}(\epsilon)=g_{\mu\nu}+\epsilon\,{h}_{\mu\nu}\,,\,\tilde{F}_{\mu\nu}(\epsilon)=F_{\mu\nu}+\epsilon\,\mathfrak{f}_{\mu\nu},
\end{eqnarray}
where $\mathfrak f_{\mu\nu}$ is related to the electromagnetic-potential perturbation $a_{\mu}$,
\begin{eqnarray}
\mathfrak{f}_{\mu\nu}=a_{\nu,\mu}-a_{\mu,\nu}\,.
\end{eqnarray}
$h_{\mu\nu}$ and $\mathfrak f_{\mu\nu}$ satisfy the linearized Einstein--Maxwell equations, whose covariant form reads
\begin{widetext}
\begin{eqnarray}
h^{\rho}_{~(\mu;\nu);\rho} -\frac{1}{2}\big(h_{\mu\nu}R+h^{\rho}_{~\rho;\mu;\nu}+{h_{\mu\nu;\rho}}^{;\rho}\big)
+\frac{1}{2}g_{\mu\nu}\big(h^{\rho\sigma}R_{\rho\sigma}-h^{\rho\sigma}_{~~;\rho;\sigma}+{h^{\rho}_{~\rho;\sigma}}^{;\sigma}\big)+\Lambda h_{\mu\nu}
\nonumber
\\
=2\big(F_{\mu\rho}{\mathfrak{f}_{\nu}}^{\rho}+\mathfrak{f}_{\mu\rho}{F_{\nu}}^{\rho}-F_{\mu\rho}F_{\nu\sigma}h^{\rho\sigma}\big)
-\frac{1}{2}h_{\mu\nu}F_{\rho\sigma}F^{\rho\sigma}+g_{\mu\nu}\big(F_{\rho\sigma}{F_{\lambda}}^{\sigma}h^{\rho\lambda}-\mathfrak{f}_{\rho\sigma}F^{\rho\sigma}\big)\,,
\label{eq-perEinstein}
\\
{\mathfrak{f}^{\mu\nu}}_{;\nu}-h^{\mu\nu}{F_{\nu\rho}}^{;\rho}-h_{\nu\rho}F^{\mu\nu;\rho}+\frac{1}{2}F^{\mu\nu}\big({h^{\rho}}_{\rho;\nu}-2{h_{\nu\rho}}^{;\rho}\big)
-\frac{1}{2}F^{\nu\rho}\big({h_{\nu\rho}}^{;\mu}+2{h^{\mu}}_{[\nu;\rho]}\big)=0\,.
\label{eq-perMaxwell}
\end{eqnarray}
\end{widetext}
The semicolon here denotes covariant differentiation with respect to the background metric $g_{\mu\nu}$, while parentheses and square brackets indicate symmetrization and antisymmetrization of indices, respectively.

To facilitate separation of variables, we expand the perturbations in spherical harmonics. Given the spherical symmetry of the black hole background, we decompose the spacetime manifold as
$\mathcal{M}=\mathcal{M}^2\times S^2$, where $\mathcal{M}^2$ is the two-dimensional $(t,r)$ submanifold and $S^2$ is the unit 2-sphere \cite{Martel:2005ir,Spiers:2023mor}. We use lowercase Latin letters $a, b, \dots$ to index tensor components on the $\{t,r\}$ submanifold $\mathcal{M}^2$, and uppercase Latin letters $A, B, \dots$ to index those on the 2-sphere $S^2$. The scalar spherical harmonics $Y^{\ell m}(\theta,\phi)$, which provide a complete basis for functions on $S^2$, are defined as the eigenfunctions satisfying
\begin{equation}
\Omega^{AB}Y^{\ell m}_{|AB}+\ell(\ell+1)Y^{\ell m}=0\,.
\label{definitionYlm}
\end{equation}
The symbol "$|$" here denotes the covariant derivative with respect to the unit sphere metric $\Omega_{AB}$, which, together with the Levi-Civita tensor $\epsilon_{AB}$ on $S^2$, is given by
\begin{eqnarray}
\Omega_{AB}=\text{diag}(1,\sin^2\theta)\,,\quad\epsilon_{\theta\phi}=-\epsilon_{\phi\theta}=\sin\theta\,.
\end{eqnarray}
With \eqref{definitionYlm}, the vector and tensor spherical harmonics can be defined as
\begin{eqnarray}
\text{Vector} : X^{\ell m}_A&=&-\epsilon_{A}^{~B}Y^{\ell m}_{|B}\,,
\\
Y^{\ell m}_A&=&Y^{\ell m}_{|A}\,.
\\
\text{Tensor}: X^{\ell m}_{AB}&=&X^{\ell m}_{(A|B)}\,,
\\
T^{\ell m}_{AB}&=&\Omega_{AB}Y^{\ell m}\,,
\\
Y^{\ell m}_{AB}&=&Y^{\ell m}_{|AB} +\tfrac{\ell(\ell+1)}{2}\Omega_{AB}Y^{\ell m}\,.
\end{eqnarray}

We expand the metric perturbation $h_{\mu\nu}$ and the electromagnetic-potential perturbation $a_{\mu}$ in spherical harmonics. Due to the spatial reflection symmetry, $(\theta,\phi)\to(\pi-\theta,\phi+\pi)$, the perturbation modes can be classified into odd parity $[(-1)^{\ell+1}]$ and even parity $[(-1)^{\ell}]$ sectors. For the odd-parity sector, the harmonic expansion takes the form
\begin{eqnarray}
h^{\rm odd}_{\mu\nu}&=&\sum_{\ell m}
\begin{pmatrix}
	0 & {h}^{\ell m}_{a}{X}^{\ell m}_{A} \\
	\text{Sym} & {h}^{\ell m}_{2}{X}^{\ell m}_{AB} \\
\end{pmatrix}\,,
\label{oddpergen}
\\
a^{\rm odd}_{\mu}&=&\sum_{\ell m}(0\,,\, B^{\ell m}{X}^{\ell m}_{A})\,,
\label{oddEMpergen}
\end{eqnarray}
whereas for the even-parity sector, the perturbations are given by
\begin{align}
h^{\rm even}_{\mu\nu}&=\sum_{\ell m}
\begin{pmatrix}
	h^{\ell m}_{ab}Y^{\ell m} & \jeven^{\ell m}_{a}{Y}^{\ell m}_{A} \\
	\rm Sym \!\!&\!\! r^2 \big(K^{\ell m}T_{AB}{+}G^{\ell m}{Y}^{\ell m}_{AB}\big) \\
\end{pmatrix},
\label{evenpergen}
\\
a^{\rm even}_{\mu}&=\sum_{\ell m}(E^{\ell m}_{a}{Y}_{\ell m}\,,\, E^{\ell m}_{2}{Y}^{\ell m}_{A})\,.
\label{evenEMpergen}
\end{align}
For brevity, the spherical harmonic indices of the expansion coefficients such as $h^{\ell m}_{a}$, $E^{\ell m}_{a}$ and \emph{et al.} will be omitted in the subsequent discussion. We will also adopt the following symbolic conventions: 
\begin{align}
h_t&\equiv{}h_0\,,\,h_r\equiv{}h_1\,,
\\
h_{tt}&\equiv{}fH_0\,,\,h_{tr}\equiv{}H_1\,,\,h_{rr}\equiv{}H_2/f\,,
\\
\jeven_t&\equiv{}\jeven_0\,,\,\jeven_r\equiv{}\jeven_1\,,
\\
E_{t}&\equiv{}E_0\,,\,E_{r}\equiv{}E_1\,.
\end{align}
The function $G$ in Eq.~\eqref{evenpergen} is the same as the equal-name $G_{RW}$ of Regge-Wheeler formulation \cite{Regge:1957td}, but $K$ is a little different, $K\to{K}_{RW}-\frac{\ell(\ell+1)}{2}G_{RW}$.

Through this decomposition, the covariant perturbation equations will be reduced to partial differential equations on the two-dimensional $(t,r)$ submanifold for each given angular quantum number $\ell$. Due to the orthogonality of spherical harmonics and parity symmetry, the odd- and even-parity modes decouple entirely at the linear perturbation level, allowing them to be analyzed separately. However, this decoupling no longer holds when a black hole carries both electric and magnetic charges. In such cases the odd- and even-parity sectors become coupled. A detailed treatment of this coupled case can be found in Ref.\cite{Pereniguez:2023wxf}. In the next section, we construct the master functions and equations by systematically extracting information from these reduced field equations without imposing gauge-fixing conditions in advance.

\section{Construction of Master Functions and Equations}
\label{construction-master-functions-and-equations}

Having reduced the four-dimensional field equations to a system of partial differential equations on the two-dimensional $(t,r)$ manifold, we now describe the construction of master functions and their associated equations. Following Ref.~\cite{Lenzi:2021wpc}, we base our approach on the following fundamental assumptions: (i) The master function is a linear combination of the perturbation variables defined on $\mathcal{M}^2$ and their first-order derivatives; (ii) the coefficients in these linear combinations depend solely on the radial coordinate $r$; (iii) the master function satisfies the canonical wave equation
\begin{equation}
\left[\frac{\partial^2}{\partial x^2}-\frac{\partial^2}{\partial t^2}-V(r)\right] \Psi(t,r)=0\,,
\label{master-equation}
\end{equation}
where the tortoise coordinate $x$ is defined via $dx/dr=f^{-1}(r)$ and the effective potential $V(r)$ remains to be determined. In what follows, we analyze the odd- and even-parity modes separately.

\subsection{Odd-parity modes}\label{odd-parity}

By the decomposition \eqref{oddpergen}-\eqref{oddEMpergen}, the odd-parity sector of the linearized Einstein--Maxwell equation \eqref{eq-perEinstein}--\eqref{eq-perMaxwell} reduces to the following component equations:
\begin{eqnarray}
&&\dot{h}_1'-h_0''+\text{LDTs}=0\,,
\label{odd-Gt}
\\
&&\ddot{h}_1-\dot{h}_0'+\text{LDTs}=0\,,
\label{odd-Gr}
\\
&&f^2{h}_2''-\ddot{h}_2+\text{LDTs}=0\,,
\label{odd-Gphi}
\\
&&f^2{B}''-\ddot{B}+\text{LDTs}=0\,,
\label{odd-B}
\end{eqnarray}
where LDTs denotes \textit{lower-order derivative terms}, that is, terms involving derivatives of a lower order than those explicitly shown. We adopted simplified notations $\phi' \equiv \partial \phi(t,r)/\partial r$ and $\dot{\phi} \equiv \partial \phi(t,r)/\partial t$. The system involves four independent functions: $h_0(t,r)$, $h_1(t,r)$, $h_2(t,r)$ and $B(t,r)$. We construct the master function $\Psi_{\text{odd}}$ as a linear combination of these functions and their first-order derivatives. In its most general form, it can be written as
\begin{align}
\Psi_{\rm odd}=C_0h_0+C_1h_1+C_2h_2+C_3\dot{h}_0+C_4{h}_0'+C_5\dot{h}_1
\nonumber
\\
+C_6{h}_1'+C_7\dot{h}_2+C_8{h}_2'+P_0B+P_1\dot{B}+P_2B'\,,
\label{psioddgen}
\end{align}
where the coefficients $C_i$ (for $i=0,1,\dots,8$) and $P_i$ (for $i=0,1,2$) depend solely on $r$. To determine these coefficients along with the potential $V(r)$, we follow the method of Ref.~\cite{Lenzi:2021wpc}. First, substituting Eq.~\eqref{psioddgen} into Eq.~\eqref{master-equation} will yield a constraint involving $h_0(t,r)$, $h_1(t,r)$, $h_2(t,r)$, $B(t,r)$, and their derivatives up to third order. Subsequently, simplifying this constraint by Eqs.~\eqref{odd-Gt}--\eqref{odd-B} and their first-order time or spatial derivatives. In doing so, we eliminate terms such as $\dddot{h}_{1}\,$, $\ddot{h}'_{1}\,$, $\dot{h}''_{1}\,$, $\dddot{h}_{2}\,$, $\ddot{h}'_{2}\,$, $\dddot{B}\,$, $\ddot{B}'\,$, $\ddot{h}_{1}\,$, $\dot{h}'_{1}\,$, $\ddot{h}_{2}\,$, $\ddot{B}\,$ and obtain
\begin{align}
\tau_{0}\,\dddot{h}_{0}+\tau_{1}\,\ddot{h}'_{0}+\tau_{2}\,\dot{h}''_{0}+\tau_{3}\,{h}'''_{0}+\tau_{4}\,{h}'''_{1}+\tau_{5}\,\ddot{h}_{0}
\nonumber
\\
+\tau_{6}\,\dot{h}'_{0}+\tau_{7}\,{h}''_{0}+\tau_{8}\,{h}''_{1}+\tau_{9}\,\dot{h}'_{2}+\tau_{10}\,{h}''_{2}+\tau_{11}\,\dot{B}'
\nonumber
\\
+\tau_{12}\,{B}''+\tau_{13}\,\dot{h}_{0}+\tau_{14}\,{h}'_{0}+\tau_{15}\,\dot{h}_{1}+\tau_{16}\,{h}'_{1}
\nonumber
\\
+\tau_{17}\,\dot{h}_{2}+\tau_{18}\,{h}'_{2}+\tau_{19}\,\dot{B}+\tau_{20}\,{B}'+\tau_{21}\,{h}_{0}
\nonumber
\\
+\tau_{22}\,{h}_{1}+\tau_{23}\,{h}_{2}+\tau_{24}\,{B}=0\,.
\label{odd-tau}
\end{align}
Here, $\tau_i$ (for $i=0,1,\cdots,24$) are functions of $C_i(r)$ (for $i=0,1,\cdots,8$), $P_i(r)$ (for $i=0,1,2$), $V(r)$, and their derivatives. Since the perturbation function $h_0(t,r)$, $h_1(t,r)$, $h_2(t,r)$, and $B(t,r)$ and their derivatives can be considered independent function basis, the condition that master function \eqref{psioddgen} satisfy the master equation implies that, all coefficient functions $\tau_i(i=0\cdots24)$ must vanish independently. This will lead to a system of algebraic-differential constraints for $C_i(r)$, $P_i(r)$, and $V(r)$.

Based on the constraints $\tau_0=\cdots=\tau_{13}=0$ and $\tau_{16}=0$, we obtain
\begin{align}
C_3(r)&=C_6(r)=0\,,\quad C_5(r)=-C_4(r)\,,
\\
C_7(r)&=-\frac{K_7}{2r}\,,\quad C_4(r)=-\frac{1}{2}\big(rC_0(r)-K_7\big)\,,
\\
C_1(r)&=\frac{K_1f(r)}{r}\,,\quad C_2(r)=\frac{K_1f(r)}{r^2},
\\
C_8(r)&=-\frac{K_1f(r)}{2r},P_1(r)={G_1},P_2(r)={G_2}f(r),
\end{align}
where $K_1$, $K_7$, $G_1$, $G_2$ are constants arising from the integration of the differential constraints. Solving the remaining constraints $\tau_{14}=0$, $\tau_{15}=0$, and $\tau_{17}=\cdots=\tau_{24}=0$, yields six independent equations
\begin{widetext}
\begin{align}
K_1\left[\frac{V(r)}{f(r)}-\bigl(\frac{\mu^2+2}{r^2}-\frac{6M}{r^3}+\frac{4Q^2}{r^4}\bigr)\right]-G_1\frac{\mu^2Q}{r^3}&=0\,,
\label{eqk1}
\\
K_1\frac{4Q}{r^3}-G_1\left[\frac{V(r)}{f(r)}-\big(\frac{\mu^2+2}{r^2}+\frac{4Q^2}{r^4}\big)\right]&=0\,,
\label{eqg1}
\end{align}
\begin{align}
{C_0'}(r)+\frac{K_7}{\mu^2} \left[\frac{V(r)}{f(r)}-\frac{f'(r)}{r}\right]-\frac{{G_2}Q}{r^3}&=0\,,
\label{eqc0p}
\\
{P_0'}(r)-2Q\left[\frac{{C_0'}(r)}{r}-\frac{{C_0}(r)}{r^2}+\frac{{K_7}}{r^3}\right]-\frac{G_2}{2}\left[\frac{V(r)}{f(r)}-\big(\frac{\mu^2+2}{r^2}+\frac{4Q^2}{r^4}\big)\right]&=0\,,
\label{eqp0p}
\\
f(r)C_0''(r)+\left[f'(r)-\frac{2f(r)}{r}\right]C_0'(r)-\left[\frac{V(r)}{f(r)}-(\frac{\mu^2+2}{r^2}-\frac{6M}{r^3}+\frac{8 Q^2}{r^4})\right]C_0(r)&
\nonumber
\\
-\frac{2Q}{r^3}P_0(r)+\frac{K_7}{r}\left[\frac{V(r)}{f(r)}-\frac{\mu^2+2f(r) }{r^2}-\frac{4Q^2}{r^4}\right]-2G_2Q\left[\frac{f'(r)}{r^3}-\frac{4 f(r)}{r^4}\right]&=0\,,
\label{eqc0pp}
\\
f(r)P_0''(r)+f'(r)P_0'(r)-\left[\frac{V(r)}{f(r)}-\frac{\mu^2+2}{r^2}\right]P_0(r)-\frac{2Q(\mu^2+2)}{r^3}C_0(r)&
\nonumber
\\
+\frac{2K_7Q(\mu^2+2)}{r^4}+\frac{G_2(\mu^2+2)}{r^2}\left[f'(r)-\frac{2f(r)}{r}\right]&=0\,,
\label{eqp0pp}
\end{align}
\end{widetext}
where $\mu^2\equiv\ell^2+\ell-2$. For charged multipoles with $Q\neq0$ and $\mu^{2}\neq0$, the solutions of Eqs.~\eqref{eqk1}--\eqref{eqp0pp} can be classified into two types. In the first type $K_1G_1\neq0$, while in the second type $K_1=0$ and $G_1=0$. 

In the first type, Eq.~\eqref{eqk1} and Eq.~\eqref{eqg1} can be considered a homogeneous linear system of $K_1$ and $G_1$. Nontrivial solution, i.e., $K_1G_1\neq0$ exists if and only if the coefficient determinant equals zero. As a result,
\begin{equation}
V_1(r)=f(r)\Big(\frac{\mu^2+2}{r^2}+\frac{p-6 M}{r^3}+\frac{4 Q^2}{r^4}\Big)\,,
\label{oddv1}
\end{equation}
where
\begin{equation}
p\equiv{}3 M\pm \sqrt{9M^2+4\mu^2Q^2}\,.
\label{p-parameter}
\end{equation}
In this case,
\begin{align}
\frac{G_1}{K_1}&=\frac{p}{\mu^2Q}\Rightarrow{}P_1(r)=G_1=\frac{K_1p}{\mu^2Q}\,,
\label{K1G1cond1}
\\
C_1(r)&=rC_2(r)=-2C_8(r)=K_1\frac{f(r)}{r}\,.
\label{C1C2C8}
\end{align}
The plus and minus signs in \eqref{p-parameter} correspond to two independent solution branches, which we treat simultaneously. As long as $K_1$, $G_1$ and $V(r)$ are determined in this way, the remaining variables $C_0(r)$, $P_0(r)$, $K_7$, and $G_2$ will be determined through Eqs.~\eqref{eqc0p}--\eqref{eqp0pp}. A direct analysis shows that these equations are consistent only if $K_7=G_2=0$. Under this condition, we obtain
\begin{align}
C_0(r)&=K_0\,,\,P_0(r)=K_0\big(\frac{2Q}{r}-\frac{p}{2 Q}\big)\,,
\label{C0P0solA}
\\
\Rightarrow{}C_5(r)&=-C_4(r)=K_0\frac{r}{2}\,,
\label{C4P5solA}
\end{align}
where $K_0$ is an arbitrary constant. 

Substituting \eqref{K1G1cond1}--\eqref{C4P5solA} into Eq.~\eqref{psioddgen}, we will get final expressions of $\Psi_\mathrm{odd}$. Since two independent constants $K_0$ and $K_1$ appears here, the general master function $\Psi_{\rm odd}(t,r)$ is a linear combination of two fundamental master functions $\Psi_0(t,r)$ and $\Psi_1(t,r)$, namely
\begin{widetext}
\begin{eqnarray}
&&\hspace{-5mm}\Psi_{\rm odd}^{K_1G_1\neq0}(t,r)=K_0\big[h_0(t,r)+\frac{r}{2}(\dot{h}_1(t,r)-h_0'(t,r))+\big(\frac{2Q}{r}-\frac{p}{2Q}\big)B(t,r)\big]
\label{psiodd0}
\\
&&\hspace{15mm}+K_1\big[\frac{f(r)}{r}\Big(h_1(t,r)+\frac{h_2(t,r)}{r}-\frac{h_2'(t,r)}{2}\Big)+\frac{p}{\mu^2 Q}\dot{B}(t,r)\big]
\nonumber
\\
&&\hspace{-5mm}\equiv{}K_0\Psi_0(t,r)+K_1\Psi_1(t,r)\,.
\label{psioddS}
\end{eqnarray}
\end{widetext}
However, using Eqs.~\eqref{odd-Gt}--\eqref{odd-B}, we can prove that $\Psi_1$ is given by the first-order time derivative of $\Psi_0$,
\begin{equation}
\Psi_1(t,r)=-\frac{2}{\mu^2}\dot{\Psi}_0(t,r)\,.
\end{equation}
It is worth noting that the master function $\Psi_\mathrm{odd}$ in Eq.~\eqref{psiodd0} contains terms proportional to $p/Q$, which may appear singular in the limit $Q\to0$. However, for the branch
$p{=}3M{-}\sqrt{9M^2{+}4\mu^2Q^2}$, the ratio $p/Q$ remains finite. For the other branch $p{=}3M{+}\sqrt{9M^2{+}4\mu^2Q^2}$, although $p/Q$ diverges, the combinations $K_0p/Q$ and $K_1p/Q$ remains finite. Hence, the master function is well defined in both cases. We will revisit this point in detail at the end of this section.

In the second type, i.e. when $K_1=G_1=0$, Eqs.~\eqref{eqk1}--\eqref{eqg1} no longer constrain $V(r)$. Instead, we need determine $K_7$, $G_2$, $C_0(r)$, $P_0(r)$, and $V(r)$ from Eqs.~\eqref{eqc0p}--\eqref{eqp0pp}. Directly solving these equations is challenging due to nonlinear terms such as $C_0(r)V(r)$. Our strategy is to differentiate Eqs.~\eqref{eqc0p}--\eqref{eqp0p} with respect to $r$ and then combine them with Eqs.~\eqref{eqc0p}--\eqref{eqp0pp} to form a six-dimensional linear algebraic system,
\begin{eqnarray}
\mathcal{A}\cdot \mathcal{X}=\mathcal{B}\,,
\label{oddLeq}
\end{eqnarray}
where
\begin{eqnarray}
\mathcal{X}\equiv\left\{C_0(r),C_0'(r),C_0''(r),P_0(r),P_0'(r),P_0''(r)\right\}^{T}.
\end{eqnarray}
The coefficient matrix $\mathcal{A}$ and the inhomogeneous vector $\mathcal{B}$ defined as follows
\begin{eqnarray}
\mathcal{A}=
\begin{pmatrix}
	0 & 1 & 0 & 0 & 0 & 0 \\
	\frac{2Q}{r^2} & -\frac{2Q}{r} & 0 & 0 & 1 & 0 \\
	* & f'-\frac{2f}{r} & f & -\frac{2Q}{r^3} & 0 & 0 \\
	-\frac{2Q(\mu^2+2)}{r^3} & 0 & 0 & * & f' & f \\
	0 & 0 & 1 & 0 & 0 & 0 \\
	-\frac{4Q}{r^3} & \frac{4Q}{r^2} & -\frac{2Q}{r} & 0 & 0 & 1
\end{pmatrix}\,,
\\
\mathcal{B}=
K_7
\begin{pmatrix}
	* \\
	\frac{2Q}{r^3} \\
	* \\
	-\frac{2Q(\mu^2+2)}{r^4} \\
	* \\
	-\frac{6Q}{r^4}
\end{pmatrix}
+G_2
\begin{pmatrix}
	\frac{Q}{r^3} \\
	* \\
	\frac{2Qf'}{r^3}-\frac{8Q f}{r^4} \\
	-(\mu^2+2)\Big(\frac{f}{r^2}\Big)' \\
	-\frac{3Q}{r^4} \\
	*
\end{pmatrix}\,.
\end{eqnarray}
Here, the asterisk denotes terms related with $V(r)$. As an algebraic system, Eq.~\eqref{oddLeq} has two solution branches. Branch 1 occurs when $\det\mathcal{A}=0$ and $\operatorname{rank}(\mathcal{A})=\operatorname{rank}(\mathcal{A}\mid\mathcal{B})$, which implies that
\begin{align}
V_2(r)&=f(r)\Big(\frac{\mu^2+2}{r^2}+\frac{p-6 M}{r^3}+\frac{4 Q^2}{r^4}\Big)\,,
\\
{G_2}&=\frac{p}{\mu^2Q}{K_7}\,.
\label{K7G2Branch1}
\end{align}
$K_7=0$ here is allowable. In this branch, Eq.~\eqref{oddLeq} admits an infinite set of solutions. However, upon verifying the differential consistency of each component in $\mathcal{X}$, we find that $K_7=G_2=0$ and $C_0(r)$, $P_0(r)$ reduce to the expressions given in Eq.~\eqref{C0P0solA}. Since $V_2(r)=V_1(r)$ and $K_1=0$, the resulting master function is simply a special case of Eq.~\eqref{psioddS}, rather than a genuinely new branch.

Branch 2 is defined by $\det\mathcal{A}\neq0$, so that the formal solution of the system is
\begin{equation}
\mathcal{X}=\mathcal{A}^{-1}\mathcal{B}\,,
\label{oddXformalsol}
\end{equation}
although the function $V(r)$ remains undetermined. We then utilize the differential consistency among the components of $\mathcal{X}$ to extract constraints on $V(r)$. Specifically, two independent differential constraints are obtainable
\begin{eqnarray}
(\partial_r\mathcal{X}_1-\mathcal{X}_2)[V,V',V'']=0\,,
\label{O3consis1}
\\
(\partial_r\mathcal{X}_4-\mathcal{X}_5)[V,V',V'']=0\,,
\label{O3consis2}
\end{eqnarray}
each being a nonlinear function of $V$, $V'$, and $V''$. It is highly impossible to have a single function $V(r)$ satisfy two manifestly distinct nonlinear differential equations. However, we find that, when $K_7$ and $G_2$ are related by
\begin{equation}
\frac{G_2}{K_7}=\frac{p}{\mu^2Q}\,,
\label{K7G2relation}
\end{equation}
this become  possible. Different from \eqref{K7G2Branch1}, $K_7=0$ here is not allowed.  Under this condition, Eq.~\eqref{O3consis1} and Eq.~\eqref{O3consis2} collapse into a single second-order nonlinear ordinary differential equation for $V(r)$, 
\begin{equation}
f(r)\Big[f(r)\frac{V'(r)+{V_1}'(r)}{V(r)-{V_1}(r)}\Big]'-\big[V(r)-{V_1}(r)\big]=0\,.
\label{eq-oddv}
\end{equation}
In principle, solutions to Eq.~\eqref{eq-oddv} always exist. For example, the function given in Eq.~\eqref{evenv3} in Sec. \ref{even-parity} provides an example, which will be further discussed in Sec. \ref{Darboux-transformation}. As long as an allowable $V(r)$ is obtained from Eq.~\eqref{eq-oddv}, the algebraic equations \eqref{oddXformalsol} can be solved routinely, and the master function can be constructed as follows
\begin{align}
\Psi_{\rm odd}^{K_1G_1=0}(t,r)&=C_0(r)\Big[h_0(t,r)-\frac{r}{2}\big(h_0'(t,r)-\dot{h}_1(t,r)\big)\Big]
\nonumber
\\
+&P_0(r)B(t,r)+K_7\Big\{\frac{1}{2}\big[h_0'(t,r)-\dot{h}_1(t,r)\big]
\nonumber
\\
&\quad-\frac{\dot{h}_2(t,r)}{2r}+\frac{p}{\mu^2Q}f(r)B'(t,r)\Big\}\,.
\label{psioddD}
\end{align}
The functions $C_0(r)$ and $P_0(r)$ here are obtained from the first and fourth components of Eq.~\eqref{oddXformalsol} and further simplified using Eq.~\eqref{K7G2relation} and Eq.~\eqref{eq-oddv}
\begin{align}
C_0(r)&=-\frac{K_7}{\mu^2}\left[f(r)\frac{V'(r)+{V_1}'(r)}{V(r)-{V_1}(r)}-\frac{2f(r)+\mu^2}{r}\right]\,,
\\
P_0(r)&=\frac{K_7f(r)}{\mu^2}\left[\Big(\frac{p}{2Q}-\frac{2Q}{r}\Big)\frac{V'(r)+{V_1}'(r)}{V(r)-{V_1}(r)}+\frac{4Q}{r^2}\right]\,.
\end{align}
It is worth emphasizing that the definitions of $p$ in both Eqs.~\eqref{K7G2Branch1} and~\eqref{K7G2relation} are the same as Eq.~\eqref{p-parameter}. So the expressions in Eq.~\eqref{psioddD} represent two master functions.

Our analysis above shows that the full algebraic-differential constraints $\tau_i=0$ (for $i=0,1,\cdots,24$) admits four branch of solutions, which ultimately lead to four distinct formulations of the master function and its associated differential equation. Two of these branches arise when $V=V_1$, corresponding to the plus/minus choices in Eq.~\eqref{oddv1}. These branches reflect two physical degrees of freedom for odd-parity perturbations, with the master function given by Eq.~\eqref{psioddS} and with $K_1$, $G_1$ given by Eq.~\eqref{K1G1cond1} and $K_7=G_2=0$. Note that the case $K_1=G_1=0$ can also be included in these branches. Since the case $V=V_1$ is directly related to the classic results of Zerilli and Moncrief on RN black hole perturbations \cite{Zerilli:1974ai,Moncrief:1974gw,Moncrief:1974ng}, extended to cases with a cosmological constant in \cite{Mellor:1989ac,Berti:2003ud}, we refer to these two plus/minus branches collectively as \emph{standard branches}. Conversely, when $V\neq V_1$, two additional branches emerge in which $K_1=G_1=0$ and but $K_7G_2\neq0$ and satisfy Eq.~\eqref{K7G2relation}. In this scenario, the potential $V(r)$ is determined by Eq.~\eqref{eq-oddv}, and the master function, as given by Eq.~\eqref{psioddD}, depends on the form $V(r)$. Following Ref.~\cite{Lenzi:2021njy}, these branches are collectively termed \emph{Darboux branches}.

This concludes our construction of the master functions and equations in the odd-parity sector.

\subsection{Even-parity modes}\label{even-parity}

Using the decomposition formulas \eqref{evenpergen}--\eqref{evenEMpergen}, the even-parity components of Eq.~\eqref{eq-perEinstein} have the following structure:
\begin{align}
&{K}''+\text{LDTs}=0\,,
\label{eveneqa11}
\\
&\dot{K}'+\text{LDTs}=0\,,
\label{eveneqa12}
\\
&\ddot{K}+\text{LDTs}=0\,,
\label{eveneqa22}
\\
&\dot{\jeven}_1'-\jeven_0''+\text{LDTs}=0\,,
\label{eveneqa13}
\\
&\ddot{\jeven}_1-\dot{\jeven}_0'+\text{LDTs}=0\,,
\label{eveneqa23}
\\
&f^2{G}''-\ddot{G}+\text{LDTs}=0\,,
\label{eveneqa33}
\\
f^2K''-\ddot{K}&-f^2H_0''+2f\dot{H}_1'+\ddot{H}_2+\text{LDTs}=0\,,
\label{eveneqa44}
\end{align}
and those from Eq.~\eqref{eq-perMaxwell} have the form
\begin{eqnarray}
&&E_0''-\dot{E}_1'+\text{LDTs}=0\,,
\label{eveneqf1}
\\
&&\dot{E}_0'-\ddot{E}_1+\text{LDTs}=0\,,
\label{eveneqf2}
\\
&&{f^2}E_2''-\ddot{E}_2+\text{LDTs}=0\,.
\label{eveneqf3}
\end{eqnarray}
In this sector, we have ten independent functions: $H_0(t,r)$, $H_1(t,r)$, $H_2(t,r)$, $\jeven_0(t,r)$, $\jeven_1(t,r)$, $K(t,r)$, $G(t,r)$, $E_0(t,r)$, $E_1(t,r)$, and $E_2(t,r)$. The master function will be constructed as a linear combination of these functions and their first-order derivatives. The most general form can be written as
\begin{widetext}
\begin{eqnarray}
\Psi_{\rm even}&=&C_0\,{H_0}+C_1\,{H_1}+C_2\,{H_2}+C_3\,{\jeven_0}+C_4\,{\jeven_1}+C_5\,{K}+C_6\,{G}+C_7\,{\dot{H}_0}+C_8\,{H_0'}+C_9\,{\dot{H}_1}
\nonumber
\\
&+&C_{10}\,{H_1'}+C_{11}\,{\dot{H}_2}+C_{12}\,{H_2'}+C_{13}\,{\dot{\jeven}_0}+C_{14}\,{\jeven_0'}+C_{15}\,{\dot{\jeven}_1}+C_{16}\,{\jeven_1'}+C_{17}\,{\dot{K}}+C_{18}\,{K'}+C_{19}\,{\dot{G}}
\nonumber
\\
&+&C_{20}\,{G'}+P_0\,E_0+P_1\,E_1+P_2\,E_2+P_3\,\dot{E}_0+P_4\,{E}_0'+P_5\,\dot{E}_1+P_6\,{E}_1'+P_7\,\dot{E}_2+P_8\,{E}_2'\,,
\label{evenPsigen}
\end{eqnarray}
\end{widetext}
where coefficients $C_i$ (for $i=0,1,\dots,20$) and $P_i$ (for $i=0,1,\dots,8$) depend solely on $r$. By substituting this combination into Eq.~\eqref{master-equation} and simplifying the resultant constraint using Eqs.~\eqref{eveneqa11}--\eqref{eveneqf3} (and their first-order derivatives), we systematically eliminate terms in the following order: step 1, the third-order derivative terms, $\dddot{H}_2$, $\ddot{H}'_2$, $\dddot{\jeven}_1$, $\ddot{\jeven}_1'$, $\dot{\jeven}''_1$, $\dddot{K}$, $\ddot{K}'$, $\dot{K}''$, ${K}'''$, $\dddot{G}$, $\ddot{G}'$, $\dddot{E}_1$, $\ddot{E}_1'$, $\dot{E}_1''$, $\dddot{E}_2$, $\ddot{E}'_2$; step 2, the second-order derivative terms, $\ddot{H}_2$, $\ddot{\jeven}_1$, $\dot{\jeven}'_1$, $\ddot{K}$, $\dot{K}'$, ${K}''$, $\ddot{G}$, $\ddot{E}_1$, $\dot{E}_1'$, $\ddot{E}_2$. After this elimination, the simplified form is given by
\begin{widetext}
\begin{eqnarray}
\tau_{0}\,\dddot{H}_{0}+\tau_{1}\,\ddot{H}'_{0}+\tau_{2}\,\dot{H}''_{0}+\tau_{3}\,{H}'''_{0}+\tau_{4}\,\dddot{H}_{1}+\tau_{5}\,\ddot{H}'_{1}+\tau_{6}\,\dot{H}''_{1}+\tau_{7}\,{H}'''_{1}+\tau_{8}\,\dot{H}''_{2}+\tau_{9}\,{H}'''_{2}
\nonumber
\\
+\tau_{10}\,\dddot{\jeven}_{0}+\tau_{11}\,\ddot{\jeven}'_{0}+\tau_{12}\,\dot{\jeven}''_{0}+\tau_{13}\,{\jeven}'''_{0}+\tau_{14}\,{\jeven}'''_{1}+\tau_{15}\,\dddot{E}_{0}+\tau_{16}\,\ddot{E}'_{0}+\tau_{17}\,\dot{E}''_{0}+\tau_{18}\,{E}'''_{0}+\tau_{19}\,{E}'''_{1}
\nonumber
\\
+\tau_{20}\,\ddot{H}_{0}+\tau_{21}\,\dot{H}'_{0}+\tau_{22}\,{H}''_{0}+\tau_{23}\,\ddot{H}_{1}+\tau_{24}\,\dot{H}'_{1}+\tau_{25}\,{H}''_{1}+\tau_{26}\,\dot{H}'_{2}+\tau_{27}\,{H}''_{2}+\tau_{28}\,\ddot{\jeven}_{0}+\tau_{29}\,\dot{\jeven}'_{0}
\nonumber
\\
+\tau_{30}\,{\jeven}''_{0}+\tau_{31}\,{\jeven}''_{1}+\tau_{32}\,\dot{G}+\tau_{33}\,{G}''+\tau_{34}\,\ddot{E}_{0}+\tau_{35}\,\dot{E}'_{0}+\tau_{36}\,{E}''_{0}+\tau_{37}\,{E}''_{1}+\tau_{38}\,\dot{E}'_{2}+\tau_{39}\,{E}''_{2}
\nonumber
\\
+\tau_{40}\,\dot{H}_{0}+\tau_{41}\,{H}'_{0}+\tau_{42}\,\dot{H}_{1}+\tau_{43}\,{H}'_{1}+\tau_{44}\,\dot{H}_{2}+\tau_{45}\,{H}'_{2}+\tau_{46}\,\dot{\jeven}_{0}+\tau_{47}\,{\jeven}'_{0}+\tau_{48}\,\dot{\jeven}_{1}+\tau_{49}\,{\jeven}'_{1}
\nonumber
\\
+\tau_{50}\,\dot{K}+\tau_{51}\,{K}'+\tau_{52}\,\dot{G}+\tau_{53}\,{G}'+\tau_{54}\,\dot{E}_{0}+\tau_{55}\,{E}'_{0}+\tau_{56}\,\dot{E}_{1}+\tau_{57}\,{E}'_{1}+\tau_{58}\,\dot{E}_{2}+\tau_{59}\,{E}'_{2}
\nonumber
\\
+\tau_{60}\,{H}_{0}+\tau_{61}\,{H}_{1}+\tau_{62}\,{H}_2+\tau_{63}\,{\jeven}_{0}+\tau_{64}\,{\jeven}_{1}+\tau_{65}\,{K}+\tau_{66}\,{G}+\tau_{67}\,{E}_{0}+\tau_{68}\,{E}_{1}+\tau_{69}\,{E}_{2}=0\,.
\label{even-tau}
\end{eqnarray}
\end{widetext}
Here, $\tau_i$ (for $i=0,1,2,\cdots$) are combinations of the functions $C_i(r)$ (for $i=0,1,\cdots,20$), $P_i(r)$ (for $i=0,1,\cdots,8$), $V(r)$, and their derivatives. For the master equation to hold, the independence of each term in \eqref{even-tau} implies that $\tau_i=0$ for all $i=0,1,\cdots,69$. This yields a system of algebraic-differential constraints for $C_i(r)$, $P_i(r)$, and $V(r)$.

Based on the constraints $\tau_0=\cdots=\tau_{19}=0$, we can directly derive
\begin{align}
&C_7(r)=C_8(r)=C_9(r)=C_{10}(r)=C_{11}(r)=0\,,
\label{expc7}
\\
&C_{12}(r)=C_{13}(r)=C_{16}(r)=P_3(r)=P_6(r)=0\,,
\\
&C_{15}(r)=-C_{14}(r)\,,\quad P_5(r)=-P_4(r)\,.
\end{align}
After this operation, from $\tau_{20}=\cdots=\tau_{39}=0$, we can derive the following results:
\begin{align}
&C_3(r)=\frac{K_3}{r}+\frac{2f(r)-(\mu^2+2)}{rf(r)}C_1(r)\,,
\label{expc3}
\\
&C_4(r)=\frac{K_4f(r)}{r}-(\mu^2+2)\frac{C_0(r)+C_2(r)}{r}\,,
\\
&C_{14}(r)=-C_1(r)\,,\quad C_{17}(r)=-\frac{r}{f(r)}C_1(r)\,,
\\
&C_{18}(r)=-r\bigl[C_0(r)+C_2(r)\bigr]\,,\, C_{19}(r)=-\frac{K_3r}{2},
\\
&C_{20}(r)=-\frac{K_4}{2}rf(r)\,,\quad C_0(r)=\frac{QP_4(r)}{2r^2}\,,
\\
&P_0(r)=G_0\,,\quad P_1(r)=G_1f(r)\,,
\\
&P_7(r)=-G_0\,,\quad P_8(r)=-G_1f(r)\,.
\label{expp7}
\end{align}
The integration constants $K_3$, $K_4$, $G_0$, and $G_1$ will be determined by further reasoning. To proceed, we substitute the results of Eqs.~\eqref{expc7}--\eqref{expp7} into the constraints $\tau_{40}=\cdots=\tau_{69}=0$. At this stage, the constraint $\tau_{54}=0$ will yield
\begin{align}
P_2(r)=0\,.
\label{expp2}
\end{align}
While the constraints $\tau_{40}=0$, $\tau_{46}=0$, and $\tau_{41}=0$ will lead to
\begin{align}
C_1(r)=&\frac{({G_1}Q+{K_3}r)f(r)}{r\lambda (r)}\,,\,C_6(r)=\frac{\mu^2+2}{2}C_5(r)\,,
\\
C_5(r)=&\left[\frac{\lambda(r)}{4r^2f(r)}+\frac{1}{r^2}\right]Q{P_4}(r)+\frac{\lambda(r)}{2f(r)}{C_2}(r)
\nonumber
\\
&-\frac{{G_0}Q}{2r}-\frac{{K_4}}{2}\,.
\label{expc5}
\end{align}
In these formulas, the auxiliary function $\lambda(r)$ is defined as
\begin{equation}
\lambda(r)=\mu^2-\Lambda r^2+3-3f(r)-\frac{Q^2}{r^2}.
\end{equation}
In addition, we note that the constraints $\tau_{42}=0$, $\tau_{43}=0$, $\tau_{44}=0$, $\tau_{45}=0$, $\tau_{49}=0$, $\tau_{57}=0$, and $\tau_{69}=0$ will reproduce results the same as those in Eqs.~\eqref{expp2}--\eqref{expc5}, thereby forming consistency check for the intermediate expression. The remaining constraints (including $\tau_{47}$, $\tau_{48}$, $\tau_{50}$--$\tau_{53}$, $\tau_{55}$--$\tau_{56}$, and $\tau_{58}$--$\tau_{68}$) are not mutually independent. Most are simple multiple of some others, and the rest are linear combinations of others. After removing these linear and multiplicative redundancies, only six equations remain independent. The first two involve the constants $K_3$, $G_1$
\begin{align}
&K_3\left[\frac{V(r)}{f(r)}-U(r)-\big(\mu^2-\lambda(r)\big)W(r)\right]
\nonumber
\\
&\hspace{40mm}-G_1\frac{\mu^2Q}{r}W(r)=0\,,
\label{evenk3}
\\
&G_1\left[\frac{V(r)}{f(r)}-U(r)-\frac{4Q^2}{r^2} W(r)\right]-K_3\frac{4Q}{r}W(r)=0\,,
\label{eveng1}
\end{align}
where the functions $U(r)$ and $W(r)$ are given by
\begin{align}
U(r)=&\frac{\mu^2+2}{r^2}+\frac{8Q^2f(r)}{r^4\lambda(r)}\,,
\label{evenfunU}
\\
W(r)=&\frac{2 (\mu^2+2-f(r))}{r^2 \lambda(r)}+\frac{2f(r)}{r^2\lambda (r)^2}\left(\mu^2+\frac{4Q^2}{r^2}\right)
\nonumber
\\
&-\frac{1}{r^2}\,.
\label{evenfunW}
\end{align}
The latter four involve the constants $G_0$, $K_4$
\begin{widetext}
\begin{align}
\frac{(\mu^2+2)}{r^2}\left[{P'_4}(r)-\frac{2}{r}P_4(r)\right]-\frac{G_0}{2}\left[\frac{V(r)}{f(r)}-\frac{\mu^2+2}{r^2}-\frac{4Q^2}{r^4}\right]+K_4\frac{2Q}{r^3}=0&\,,
\label{eqp4p}
\\
\frac{(\mu^2+2)Q}{2}\left\{\bigl[4f(r)+\lambda(r)\bigr]{P'_4}(r)-\left[\frac{12f(r)}{r}+\left(\frac{f'(r)}{f(r)}+\frac{3}{r}\right)\lambda(r)-\lambda'(r)\right]{P_4}(r)\right\}+{G_0}(\mu^2+4)Qf(r)&
\nonumber
\\
+(\mu^2+2)r^2\lambda(r)\left[{C'_2}(r)+\left(\frac{\lambda'(r)}{\lambda(r)}-\frac{f'(r)}{f(r)}-\frac{1}{r}\right){C_2}(r)\right]+{K_4}r^3f(r)\left[\frac{V(r)}{f(r)}-\frac{\mu^2-\lambda(r)}{r^2}\right]=0&\,,
\label{eqc2p}
\\
Qf(r){P''_4}(r)+\frac{Q^2+r^2-\Lambda r^4-5r^2 f(r)}{r^3}Q{P'_4}(r)-\left[\frac{V(r)}{f(r)}-\frac{8f(r)+2\Lambda r^2+\mu^2}{r^2}+\frac{Q^2}{r^4}\left(\frac{\lambda (r)}{f(r)}+8\right)\right]Q{P_4}(r)&
\nonumber
\\
+\frac{4Q^2}{r}{C'_2}(r)-\frac{2Q^2}{r^2}\left[\frac{\lambda (r)}{f(r)}+6\right]{C_2}(r)+{G_0}Q\frac{Q^2+r^2-\Lambda{}r^4-3r^2f(r)}{r^3}+{K_4}\frac{2Q^2}{r^2}=0&\,,
\label{eqp4pp}
\\
f(r){C_2''}(r)-\left[\frac{4 f(r)+\lambda (r)}{r}+\frac{2 Q^2}{r^3}\right]{C_2'}(r)-\left[\frac{V(r)}{f(r)}-\frac{\lambda (r) \left(1-\Lambda  r^2\right)}{r^2f(r)}-\frac{\mu^2-2\Lambda  r^2+6}{r^2}\right]{C_2}(r)-{K_4}\Bigl[\frac{f(r)}{r^2}+&\frac{Q^2}{r^4}\Bigr]
\nonumber
\\
+\frac{2\Lambda r^2-\mu^2-4}{2 r^3}QP_4'(r) +\left[\frac{\lambda (r)(1-\Lambda  r^2)}{2r^4f(r)}+\frac{\mu^2-3\Lambda  r^2+6}{r^4}\right]Q{P_4}(r)-{G_0}Q\left[\frac{f'(r)}{2 r^2}+\frac{Q^2}{r^5}\right]=0&\,.
\label{eqc2pp}
\end{align}
\end{widetext}
It can be shown that Eqs.~\eqref{evenk3}--\eqref{eveng1} and \eqref{eqp4p}--\eqref{eqc2pp} share the same structure as their odd-parity counterparts, Eqs.~\eqref{eqk1}--\eqref{eqp0pp}. Consequently, we adopt the same solution strategy. In particular, to solve the nonlinear system \eqref{eqp4p}--\eqref{eqc2pp}, we recast it into a complete six-dimensional linear algebraic system $\mathcal{A}\!\cdot\!\mathcal{X}=\mathcal{B}$, where
\begin{equation}
\mathcal{X}\equiv\left\{P_4(r),P_4'(r),P_4''(r),C_2(r),C_2'(r),C_2''(r)\right\}^{T}\,.
\end{equation}
The consistency condition of this linear system will then be analyzed to yield relations necessary to fix the unknowns. For brevity, we omit the detailed derivation of the related equations and directly summarize the undetermined quantities in the algebraic-differential constraints $\tau_i=0$ (for $i=0,1,\cdots,69$) before presenting the master function.

Depending on differences between the effective potentials, the solution space of the algebraic-differential system can be partitioned into \emph{standard branches} and \emph{Darboux branches}. Similar to the odd parity case, we also have definitions for the parameter $p=3M\pm\sqrt{9M^2+4\mu^2Q^2}$. Within each standard branch, and continuing Eqs.~\eqref{expc7}--\eqref{expc5}, the remaining quantities can be solved to yield
\begin{align}
&{V_3}(r)=f(r)\Big[U(r)+\Big(\frac{\mu^2r+p}{r}-\lambda(r)\Big)W(r)\Big]\,,
\label{evenv3}
\\
&{G_1}=\frac{p}{\mu^2Q}{K_3}\,,\,G_0=K_4=0\,,\,P_4(r)=-\frac{{K_2}p}{2\mu^2Q} r^2\,,
\label{eqevenp4}
\\
&C_2(r)={K_2}\left(\frac{(\mu^2 r+p)f(r)}{\mu^2 \lambda (r)}+\frac{p}{4 \mu^2 }\right)\,,
\label{eqevenc2}
\end{align}
where $K_2$ and $K_3$ are arbitrary constants. Substituting Eqs.~\eqref{eqevenp4}--\eqref{eqevenc2} into Eqs.~\eqref{expc3}--\eqref{expc5}, combining with Eqs.~\eqref{expc7}--\eqref{expc5}, through Eq.~\eqref{evenPsigen} we can get the master function for this branch,
\begin{equation}
\Psi_{\rm even}(t,r)=K_2\Psi_2(t,r)+K_3\Psi_3(t,r)\,,
\label{psievenS}
\end{equation}
where
\begin{widetext}
\begin{eqnarray}
\Psi_2(t,r)&=&\frac{r}{2} K(t,r)+\frac{\mu^2+2}{4}r G(t,r)-\frac{(\mu^2 r+p)f(r)}{\mu^2r\lambda (r)}\left[(\mu^2+2)\,{\jeven}_1(t,r)+r^2K'(t,r)\right]
\nonumber
\\
&-&\frac{p}{4\mu^2}{H_0}(t,r)+\left[\frac{(\mu^2 r+p)f(r)}{\mu^2 \lambda (r)}+\frac{p}{4\mu^2}\right]H_2(t,r)-\frac{pr^2}{2\mu^2Q}\left[E_0'(t,r)-\dot{E}_1(t,r)\right]\,,
\label{evenpsi2}
\\
\Psi_3(t,r)&=&\left[\frac{1}{r}-\frac{(\mu^2r+p)\big(\mu^2+2-2f(r)\big)}{\mu^2 r^2\lambda (r)}\right]{{\jeven}_0}(t,r)-\frac{\mu^2 r+p}{\mu^2 \lambda(r)}\dot{K}(t,r)-\frac{r}{2}\dot{G}(t,r)
\nonumber
\\
&+&\frac{(\mu^2 r+p)f(r)}{\mu^2 r\lambda(r)}\left[{H_1}(t,r)-{{\jeven}_0'}(t,r)+\dot{\jeven}_1(t,r)\right]+\frac{pf(r)}{\mu^2Q}\left[E_1(t,r)-E_2'(t,r)\right]\,.
\label{evenpsi3}
\end{eqnarray}
\end{widetext}
Using Eqs.~\eqref{eveneqa11}--\eqref{eveneqf3}, it can be shown that the master functions $\Psi_2(t,r)$ and $\Psi_3(t,r)$ are related by a first-order time derivative
\begin{equation}
\Psi_3(t,r)=-\frac{2}{\mu^2+2}\dot{\Psi}_2(t,r)\,.
\end{equation}
Within each Darboux branch, the coefficients $G_1$, $K_3$, $G_0$, and $K_4$, which determine the functions in \eqref{expc3}--\eqref{expc5} are
\begin{align}
G_1=K_3=0\,,\quad G_0=\frac{p}{\mu^2Q}K_4\,,
\label{expG1K3G0K4}
\end{align}
with the potential $V(r)$, determined by the following second-order nonlinear differential equation,
\begin{equation}
f(r)\Big[f(r)\frac{V'(r)+{V_3}'(r)}{V(r)-{V_3}(r)}\Big]'-\big[V(r)-{V_3}(r)\big]=0\,.
\label{eq-evenv}
\end{equation}
The effective potential in this branch is distinct from the $V_3(r)$ in the standard branch. Complicating as it is, we note that Eq.~\eqref{oddv1} provides an example of exact solutions to this differential equation. The functions $P_4(r)$ and $C_2(r)$, which depend on the form of $V(r)$, are given by
\begin{align}
P_4(r)=&\frac{{K_4}pr^2f(r)}{\mu^2Q(\mu^2+2)}\left[\frac{1}{2}\frac{V'(r)+{V_3}'(r)}{V(r)-{V_3}(r)}+\frac{1}{r}-\frac{\mu^2r+p}{r^2\lambda(r)}\right],
\label{eqevenp4v}
\\
C_2(r)=&K_4\frac{(\mu^2 r+p)f(r)}{\mu^2 r\lambda (r)}
\nonumber
\\
&\qquad-\frac{Q}{p}\left[\frac{2(\mu^2 r+p)f(r)}{r^2\lambda(r)}+\frac{p}{2r^2}\right]{P_4}(r).
\label{expc2}
\end{align}
By similar operations as those in the standard branch, i.e. substituting Eqs.~\eqref{expG1K3G0K4} and~\eqref{expc2} into Eqs.~\eqref{expc3}--\eqref{expc5}, combining with Eqs.~\eqref{expc7}--\eqref{expc5}, we get the master function of the Darboux branch through Eq.~\eqref{evenPsigen},
\begin{widetext}
\begin{align}
\Psi_{\rm even}(t,r)&=-\frac{2\mu^2Q}{p r^2}P_4(r)\Psi_2(t,r)+K_4\Biggl\{\frac{(\mu^2r+p)f(r)}{\mu^2r\lambda(r)}\left[{H_2}(t,r)-r{K'}(t,r)\right]-\frac{rf(r)}{2}G'(t,r)
\nonumber
\\
&+f(r)\left[\frac{1}{r}-\frac{(\mu^2+2)(\mu^2r+p)}{\mu^2r^2\lambda(r)}\right]\,{\jeven}_1(t,r)+\frac{p}{\mu^2Q}\left[{E_0}(t,r)-{\dot{E}_2}(t,r)\right]\Biggr\}.
\label{psievenD}
\end{align}
\end{widetext}
$\Psi_{2}(t,r)$ here is the same as that appears in the standard branch, Eq.~\eqref{evenpsi2}. This concludes our construction of the master function and equations in the even-parity sector.

In the remainder of this section, we examine the degeneration of the effective potentials and master functions of the standard branches for both the odd- and even-parity sectors in the uncharged limit ($Q\to0$). For the minus choice, the parameter $p=3M-\sqrt{9M^2+4\mu^2Q^2}$ becomes a second-order small quantity $p\sim\mathcal{O}(Q^2)$; the potentials in Eqs.~\eqref{oddv1} and~\eqref{evenv3} reduce, respectively, to the Regge-Wheeler and Zerilli potentials; moreover, the master functions given in Eqs.~\eqref{psioddS} and~\eqref{psievenS} exactly coincide with those described in \cite{Lenzi:2021wpc}. For the plus choice $p=3M+\sqrt{9M^2+4\mu^2Q^2}$, we have $p\to6M$ as $Q\to0$. In this case, the system degenerates to that of vacuum electromagnetic perturbations: both the odd- and even-parity potentials in Eqs.~\eqref{oddv1} and~\eqref{evenv3} reduce to the effective potential for electromagnetic perturbations \cite{Cardoso:2001bb},
\begin{equation}
V^{\rm EM}(r)=f(r)\frac{\mu^2+2}{r^2}\,,
\end{equation}
and the master functions from Eqs.~\eqref{psioddS} and \eqref{psievenS} degenerate, respectively, to
\begin{eqnarray}
\Psi^{\rm EM}_{\rm odd}(t,r)&=&G_0B(t,r)+G_1\dot{B}(t,r)\,,
\label{PsioddEM}
\\
\Psi^{\rm EM}_{\rm even}(t,r)&=&{G_4}r^2\left[E_0'(t,r)-\dot{E}_1(t,r)\right]\,,
\nonumber
\\
&+&{G_1}f(r)\left[E_1(t,r)-E_2'(t,r)\right] .
\label{PsievenEM}
\end{eqnarray}
The coefficients $G_0$ and $G_1$ in Eq.~\eqref{PsioddEM} are arbitrary constants arising from rescaling the original normalization parameters $K_0$ and $K_1$, specifically through $G_0=-\frac{p}{2Q}K_0$ and $G_1=\frac{p}{\mu^2Q}K_1$ [see Eq.~\eqref{K1G1cond1}]. Similarly, the coefficients $G_4$ and $G_1$ in Eq.~\eqref{PsievenEM} are obtained via analogous rescalings.

We emphasize that this rescaling is not merely a formal manipulation, but reflects a physical necessity. As $Q \to 0$, the gravitational and electromagnetic perturbations decouple, and the potential $B(t,r)$, which enters linearly into the master function, must be rescaled in order to represent a finite-energy electromagnetic mode. The electromagnetic energy density is governed by the quadratic invariant $\mathfrak{f}_{\mu\nu} \mathfrak{f}^{\mu\nu}$, which contains terms of the form $(\partial B)^2$. Therefore, to ensure a finite energy configuration in the uncharged limit, the amplitude of $B(t,r)$ must scale proportionally to $Q$. In Eq.~\eqref{PsioddEM}, the constants $G_0$ and $G_1$ effectively incorporate this scaling, implying that \(K_0, K_1 \sim Q\).

In contrast, the unnormalized master functions such as $\Psi_0$ and $\Psi_1$ defined by Eq.~\eqref{psioddS}, which appear irrespective of the constants $K_0$ and $K_1$, do not incorporate this scaling automatically. In particular, in the plus branch, $\Psi_0$ and $\Psi_1$ contain terms proportional to $p/Q$, which diverge as $Q \to 0$. This apparent singularity is not pathological, but simply reflects a mismatch between the fixed normalization of the master function and the scaling behavior required to extract the physical electromagnetic perturbation modes. It can be resolved by redefining the master function appropriately, for instance by introducing a regularized function $\hat{\Psi} = Q \Psi$, which remains finite and captures the correct physical content in the uncharged limit.

\section{Darboux Transformation for Master Equations} \label{Darboux-transformation}

The Darboux transformation is a powerful tool in the theory of integrable systems. Through a generating function, it establishes a strict map between different potential systems, while preserving the spectral structure of the differential equation, such as the QNM frequency $\omega$ \cite{Lenzi:2021njy}. By this tool, we will uncover both a possible isospectral relation between the odd- and even-parity modes, and a connection between the standard and Darboux branches within each parity sector. We begin with the potentials given in Eqs.~\eqref{oddv1} and \eqref{evenv3} and show how the Darboux transformation acts as a bridge between these distinct potential systems.

Consider the pair of master equations governing the perturbation
\begin{equation}
\left[-\partial^2_{t}+\partial^2_{x}-V^{\text{S}/\text{D}}(r)\right] \Psi^{\text{S}/\text{D}}(t,r)=0\,,
\label{eqPsiSD}
\end{equation}
where $x$ is the tortoise coordinate defined by $dx/dr=f^{-1}(r)$ and the potentials $V^{\text{S}}(r)$ and $V^{\text{D}}(r)$ take the form
\begin{align}
&V^{\text{S}}(r)=f(r)\left[\frac{\mu^2+2}{r^2}+\frac{p-6M}{r^3}+\frac{4Q^2}{r^4}\right],
\label{VeffS}
\\
&V^{\text{D}}(r)=f(r)\left[U(r)+\left(\frac{\mu^2r+p}{r}-\lambda(r)\right)W(r)\right],
\label{VeffD}
\end{align}
with $U(r)$ and $W(r)$ given by Eqs.~\eqref{evenfunU} and \eqref{evenfunW}, respectively. Note that $p=3M\pm\sqrt{9M^2+4\mu^2Q^2}$ has two choices, each corresponds to an independent pair of master equations. For definiteness we select one choice of $p$ and use it consistently in both $V^S$ and $V^D$. The core of the Darboux transformation is the generating function. For the potential system \eqref{eqPsiSD}-\eqref{VeffD}, the generating function $g(r)$ is defined by
\begin{align}
g(r)=-\frac{4Q^2f(r)}{r^3\lambda(r)}-\frac{4\mu^2Q^2f(r)}{pr^2\lambda(r)}-\frac{(\mu^2+2)p}{8Q^2}\,.
\label{generating-fun}
\end{align}
It can be verified that this form $g(r)$ satisfies the Riccati equation
\begin{align}
g_{,x}-g^2+V^{\text{S}}=\mathcal{C}\,,
\label{Riccati-eq}
\end{align}
where
\begin{equation}
\mathcal{C}=-\frac{(\mu^2+2)^2(3M p+2\mu^2Q^2)}{32Q^4}\,.
\end{equation}
Note that in the $Q \to 0$ limit, $g(r)$ and $\mathcal{C}$ become singular only on the branch $p=3M+\sqrt{9M^2+4\mu^2Q^2}$. In this case, the two effective potentials coincide, so no real transformation is needed.  On the other branch, by contrast, $g(r)$ and $\mathcal{C}$ are regular, since $p \sim -\frac{2 \mu^2 Q^2}{3 M} + \mathcal{O}(Q^4)$ and the two potentials are distinct. By the generating function $g$, we can write the mapping between two equations of Eq.~\eqref{eqPsiSD} as
\begin{align}
&\Psi^{\text{D}}=\Psi^{\text{S}}_{,x}+g\Psi^{\text{S}}\,,
\label{wavefunction-map}
\\
&V^{\text{D}}=V^{\text{S}}+2g_{,x}\,.
\label{potential-map}
\end{align}
The existence of this mapping leads to two important conclusions: (i) if $V^{\text{S}}$ and $V^{\text{D}}$ are taken as effective potentials of different parity sectors, Eq.~\eqref{wavefunction-map} establishes an isospectral correspondence between the two parity sectors. Specifically, the single-frequency solution $\Psi(t,r)=e^{i\omega t}\psi(\omega;r)$ inherits the same spectral parameter $\omega$ before and after mapping. This isospectral phenomenon was first noted by Chandrasekhar \cite{Chandrasekhar:1975nkd,Chandrasekhar:1975zza}. (ii) if $V^{\text{S}}$ and $V^{\text{D}}$ are taken as effective potentials of odd-parity but different branches, i.e. standard and Darboux branches, respectively, then Eqs.~\eqref{wavefunction-map} and \eqref{potential-map} demonstrate the full equivalence between the two-branches. For even-parity modes, we just need swap $\text{S} \leftrightarrow \text{D}$ and change $g \rightarrow -g$.

From the Riccati equation and the effective potential mapping \eqref{potential-map}, we can prove that the generating function $g$ and its derivative $g_{,x}$ can be expressed as
\begin{align}
g=\frac{(V^{\text{D}}+V^{\text{S}})_{,x}}{2\left(V^{\text{D}}-V^{\text{S}}\right)}\,,\quad g_{,x}=\frac{V^{\text{D}}-V^{\text{S}}}{2}\,.
\label{eqggx}
\end{align}
The self-consistency of these two expressions implies constraint
\begin{equation}
\Big[\frac{(V^{\text{D}}+V^{\text{S}})_{,x}}{V^{\text{D}}-V^{\text{S}}}\Big]_{,x}-\big(V^{\text{D}}-V^{\text{S}}\big)=0\,.
\label{nonlinear-constraint}
\end{equation}
Since this equation is isomorphic to the differential constraint Eq.~\eqref{eq-oddv} satisfied by the effective potential $V$ of the odd-parity perturbation, the potential $V^{\text{D}}$ here constitutes an exact implementation of $V$ there. This provides a concrete example for the conclusion (ii) above, i.e. within a given parity sector, $V^\mathrm{S}$ and $V^\mathrm{D}$ belong to two different branches.

\section{Conclusions} \label{conclusions}

In this work, we investigated the construction of master functions and equations for the perturbation of black holes coupled with electromagnetic field and cosmological constant. Without gauge-fixing constraints, our analysis is valid for any perturbative gauge. The master functions are formulated as linear combinations of the metric and electromagnetic-potential perturbations and their first-order derivatives. They satisfy wave equations with effective potentials determined by the linearized Einstein--Maxwell system.

Our results show that the master functions and their potentials exhibit a four-branch structure in both odd- and even-parity modes. In standard branches, which involve a twofold choice of the parameter $p$ [see Eq.~\eqref{p-parameter}], the effective potentials coincide with those obtained in the Zerilli-Moncrief formalism. Concretely, the master functions are precisely given by Eq.~\eqref{psioddS} for the odd-parity mode and by Eq.~\eqref{psievenS} for the even-parity mode. The newly identified Darboux branches, on the other hand, feature potentials governed by the nonlinear differential equations \eqref{eq-oddv} for the odd-parity mode and \eqref{eq-evenv} for the even-parity mode. We confirm that $V^{\text{D}}$ in Eq.~\eqref{VeffD} and $V^{\text{S}}$ in Eq.~\eqref{VeffS} are exact solutions to these equations; however, whether other globally regular solutions exist remains an open question.

Importantly, the standard branches fully encode the physical degrees of freedom in each parity sector. The Darboux branches arise from an intrinsic freedom in defining wave operators, which corresponds to a mathematical ambiguity in selecting effective potentials when decoupling the perturbation equations. This freedom is realized through the Darboux transformation, which generates distinct but physically equivalent pairs of effective potentials and master functions. As a result, the Darboux branches represent alternative mathematical descriptions rather than new dynamical degrees of freedom.

\begin{acknowledgments}

This work was supported by the Natural Science Foundation of China under Grant No. 11875082.

\end{acknowledgments}

%

\end{document}